# A first-principles study on the physical properties of two-dimensional $Nb_3Cl_8$, $Nb_3Br_8$ and $Nb_3I_8$

Bohayra Mortazavi [a,*], Xiaoying Zhuang[a,b] and Timon Rabczuk[b,**]

[a]*Chair of Computational Science and Simulation Technology, Department of Mathematics and Physics, Leibniz Universität Hannover, Appelstraße 11, 30167 Hannover, Germany.*
[b]*College of Civil Engineering, Department of Geotechnical Engineering, Tongji University, 1239 Siping Road Shanghai, China.*

**Abstract**

In a recent advance, $Nb_3Cl_8$ two-dimensional crystals with a kagome lattice and electronic topological flat bands has been experimentally fabricated (*Nano Lett. 2022, 22, 4596*). In this work motivated by the aforementioned progress, we conduct first-principles calculations to explore the structural, phonon dispersion relations, single-layer exfoliation energies and mechanical features of the $Nb_3X_8$ (X=Cl, Br, I) nanosheets. Acquired phonon dispersion relations reveal the dynamical stability of the $Nb_3X_8$ (X=Cl, Br, I) monolayers. In order to isolate single-layer crystals from bulk counterparts, we predicted exfoliation energies of 0.24, 0.27 and 0.28 $J/m^2$, for the $Nb_3Cl_8$, $Nb_3Br_8$ and $Nb_3I_8$ monolayers, respectively, which are noticeably lower than that of the graphene. We found that the $Nb_3X_8$ monolayers are relatively strong nanosheets with isotropic elasticity and anisotropic tensile strength. It is moreover shown that by increasing the atomic weight of halogen atoms in the $Nb_3X_8$ nanosheets, mechanical characteristics decline. Presented results provide a useful vision about the key physical properties of novel 2D systems of $Nb_3X_8$ (X=Cl, Br, I).

Keywords: *$Nb_3Cl_8$; mechanical; exfoliation energy; phonon dispersion*

## 1. Introduction

Since the experimental introduction of graphene [1–3] in 2004, two-dimensional (2D) crystals have been extending nonstop because of their outstanding physical properties. High surface to volume ratios in 2D systems, not only can evolve to exceptional electronic and optical features, but are also highly appealing for practical chemistry-related applications, like sensing and energy storage. During the last couple of years, several 2D crystals with interesting physical properties have been fabricated, such as $MoSi_2N_4$ [4], penta-PdPSe [5], $NbOI_2$ [6] and $NiN_2$ [7] nanosheets. The appealing prospect for the application of 2D lattices in critical technologies and their exceptional physical features, act as continuous driving force for the experimental endeavors to design and fabricate novel crystals. In line of this trend, most recently Sun *et al.* [8] succeeded in the exfoliation of the single-layer $Nb_3Cl_8$. Experimental tests and theoretical calculations confirm that the $Nb_3Cl_8$ shows topological flat bands, which was also found to offers novel opportunities to explore connection between geometry, topology and magnetism [8]. This experimental achievement also highlights the bright possibility for the exfoliation of the $Nb_3Br_8$ and $Nb_3I_8$ monolayers with similar, kagome atomic configuration. Worthy to note that $Nb_3I_8$ layered structures have been already fabricated [9] and their electronic properties have been also investigated theoretically [10–14]. Electronic and optical properties of $Nb_3Cl_8$, $Nb_3Br_8$ and $Nb_3I_8$ nanosheets have been also already studied using the density functional theory (DFT) calculations [15]. To the best of our knowledge, the stress-strain relations and failure mechanism of these 2D systems however have not been studied. In this work, our goal is to theoretically investigate the structural, phonon dispersion relations, single-layer exfoliation energies and mechanical features of the $Nb_3X_8$ (X=Cl, Br, I) nanosheets, by employing spin-polarized DFT calculations.

## 2. Computational methods

Spin-polarized DFT calculations were carried out using the *Vienna Ab-initio Simulation Package* [16,17] with the generalized gradient approximation (GGA), Perdew-Burke-Ernzerhof (PBE) exchange–correlation functional and DFT-D3 [18] vdW dispersion correction. The plane wave and self-consistent loop cutoff energies were set to 300 and $10^{-5}$ eV, respectively. In order to find geometry optimized lattices, atomic positions were relaxed using conjugate gradient algorithm until Hellman-Feynman forces drop below 0.001 eV/Å [19] with 5×5×1 and 5×5×3 Monkhorst-Pack [20] K-point grid for the monolayer and bulk systems, respectively. The stress-free structures are acquired by isotropically changing the in-plane lattice dimensions. Periodic boundary conditions were considered in all directions, with a 20

Å (vacuum distance of over 16 Å) box-size along the monolayers' thickness to avoid artificial interactions. Moment tensor potentials (MTPs) [21] were fitted to accurately evaluate the phonon dispersion relations, using the same approach as that employed in our recent work [22]. As it has been confirmed for an extensive set of 2D structures [22], MTPs can accurately reproduce DFT-based results for phononic properties. Phonon dispersions on the basis of trained MTPs were acquired in conjunction with the PHONOPY [23] code over 4×4×1 supercells, as elaborately discussed in our previous work [24].

### 3. Results and discussions

We first study the structural and bonding features of the $Nb_3Cl_8$, $Nb_3Br_8$ and $Nb_3I_8$ kagome lattices. As the representative lattice, in Fig. 1 the stress-free and energy minimized single-layer and bulk $Nb_3Cl_8$ lattices are illustrated. The hexagonal lattice constants of the $Nb_3Cl_8$, $Nb_3Br_8$ and $Nb_3I_8$ monolayers are predicted to be 6.783, 7.121 and 7.642 Å, respectively, which are considerably close to corresponding values of 6.744, 7.114 and 7.675 Å, predicted using the DFT+U method by Jiang *et al.* [15]. For the bulk lattices, the corresponding box sizes along the out-of-plane direction are predicted to be 12.320 ,13.002 and 13.998 Å, respectively (find supplementary information document for the complete crystal data). As expected, by increasing the atomic number of halogen atoms in these systems, the lattice parameter increases. Based on the Bader [25] charge analysis, stemming from the lower electronegativity of Nb atoms, it is found that they tend to by average transfer 1.53, 1.33 and 1.11 *e* to the surrounding halogen atoms in the $Nb_3Cl_8$, $Nb_3Br_8$ and $Nb_3I_8$ monolayers, respectively. The difference in the electronegativity of Nb and considered halogen atoms are also rather high, resulting in the formation of ionic interactions. This finding is also in agreement with the electron localization function (ELF) [26] section maps, shown in Fig. 1, which reveal that around the center of Nb-X bonds, the ELF shows sharp patterns, resembling the electron gas behavior, with near zero values in the direction toward the Nb atoms.

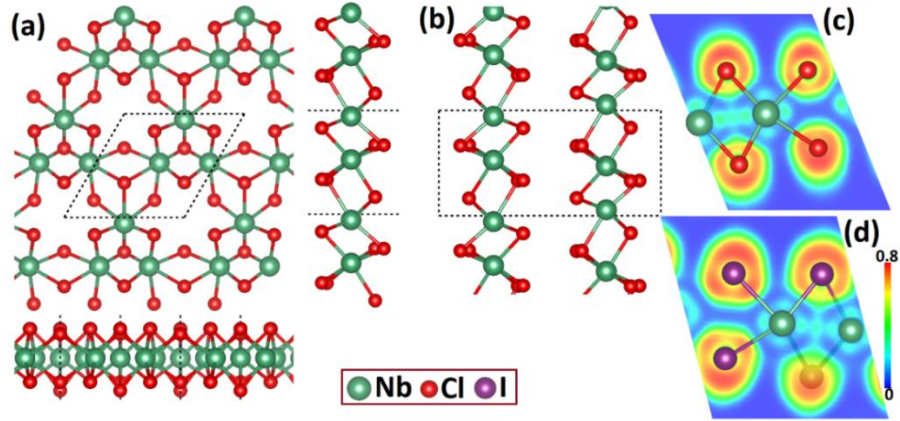

**Fig. 1**, (a) Top and side views of the $Nb_3Cl_8$ monolayer. (b) Side view of the bulk $Nb_3Cl_8$ crystal. Electron localization function (ELF) sections for the (c) $Nb_3Cl_8$ and (d) $Nb_3I_8$ monolayers.

After an effective analysis of the structural and bonding characteristics of the $Nb_3Cl_8$, $Nb_3Br_8$ and $Nb_3I_8$ kagome lattices, we next examine their dynamical stability by evaluating phonon dispersion relations. The predicted phonon dispersion along highly symmetrical points for the $Nb_3Cl_8$, $Nb_3Br_8$ and $Nb_3I_8$ monolayers are illustrated in Fig. 2, which show close agreements with those predicted by Jiang *et al.* [15]. As the first important finding, phonon modes are free of imaginary frequencies, confirming the dynamical stability of these systems. By increasing the atomic number of halogen atoms in the $Nb_3X_8$ systems, it is clear that with preserving the general form of dispersions, phonon modes show narrower frequency ranges, which indicate lower group velocity. It is also noticeable that the out-of-plane acoustic mode (ZA) in these systems almost shows no intersection with other bands, which reveal its higher lifetime. In contrast, remaining two acoustic modes and particularly all optical modes, appear with considerable intersections, stimulating the scattering and reducing the lifetime for modes with higher frequencies. It can be thus expected that likely to graphene, the ZA should be also the dominant phononic heat carrier in these system. Moreover, with generally lower group velocity and lifetime, it is expected that by increasing the atomic number of halogen atoms in $Nb_3X_8$ systems, they show lower lattice thermal conductivity.

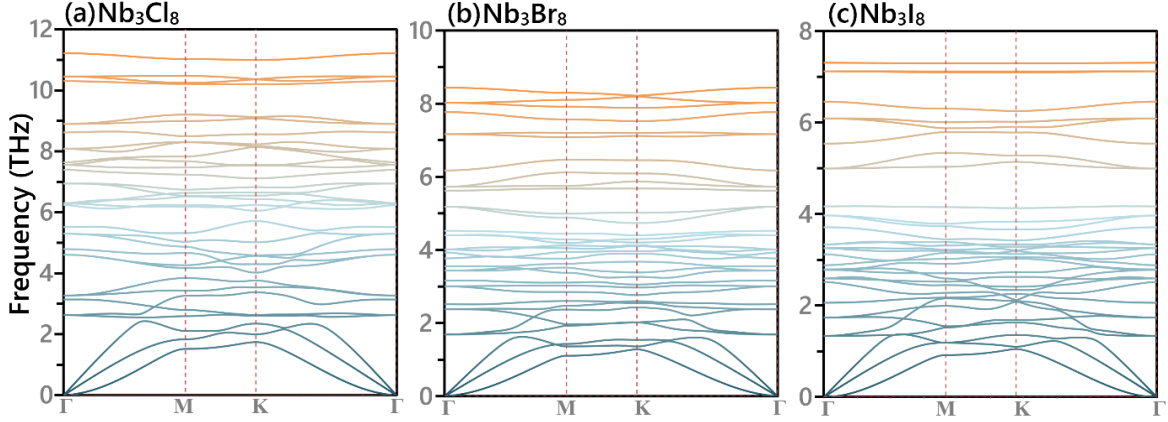

**Fig. 2**, Phonon dispersion relations of the Nb$_3$Cl$_8$, Nb$_3$Br$_8$ and Nb$_3$I$_8$ kagome monolayers.

Before analyzing the mechanical properties, it is very useful to investigate the exfoliation energy required for the isolation of the Nb$_3$Cl$_8$, Nb$_3$Br$_8$ and Nb$_3$I$_8$ kagome monolayers from their native bulk counterparts. For this purpose, we first acquired the energy minimized six-layered structures of the Nb$_3$X$_8$ systems, using the bulk structures as references for stacking pattern. In the next step, the last layer was gradually separated toward the out-of-plane vacuum direction, with a small step of 0.25 Å. The change in the energy of the systems were subsequently calculated and the cleavage energy was recorded. As shown in Fig. 3, the relative energies show sharp initial increases and later for the separation distances over around 8 Å reach converged values. According to our spin-polarized DFT-D3 simulations, the exfoliation energies of 0.24, 0.27 and 0.28 J/m$^2$ are predicted for the Nb$_3$Cl$_8$, Nb$_3$Br$_8$ and Nb$_3$I$_8$ monolayers' isolation, which are distinctly lower than that of graphene, 0.37 J/m$^2$ [27]. These findings reveal that the separate layers in these systems show weak interactions, and moreover the type of halogen atoms does not yield substantial effects on the exfoliation energy. We remind that Sun *et al.* [8] could synthesis the Nb$_3$Cl$_8$ monolayer using the mechanical exfoliation method. Taking into account this recent experimental success and our predictions for the exfoliation energies, the experimental isolation of Nb$_3$Br$_8$ and Nb$_3$I$_8$ monolayers from their bulk structures is practically feasible.

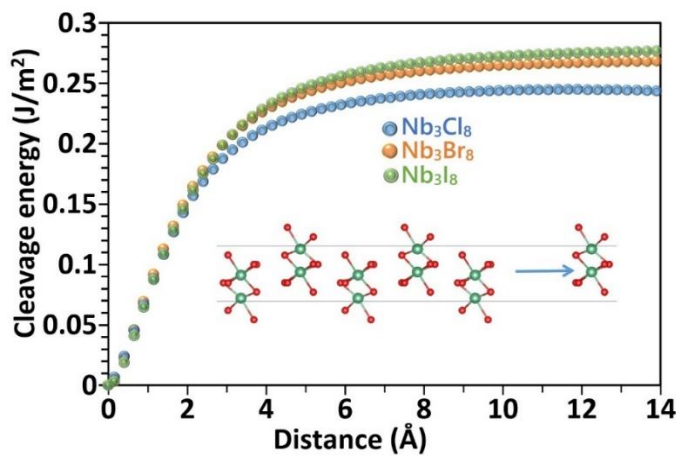

**Fig. 3**, Cleavage energy of the $Nb_3Cl_8$, $Nb_3Br_8$ and $Nb_3I_8$ monolayers as a function of separation distance.

Finally, we examine the mechanical responses by performing uniaxial tensile simulations along the armchair and zigzag directions. Uniaxial stress-strain responses of the $Nb_3X_8$ monolayers along the armchair and zigzag directions are illustrated in Fig. 4. In these results, real volumes of the deformed lattices are considered in the conversion of the stress values to the standard GPa unit. The real area of the deformed nanosheets can be easily calculated using the simulation box sizes along the in-plane directions. The real volume at every strain is thus calculated by finding the normal distance between boundary halogen atoms plus their effective vdW diameter. According to our geometry optimized bulk lattices, the effective thickness of the stress-free $Nb_3Cl_8$, $Nb_3Br_8$ and $Nb_3I_8$ monolayers are 6.16, 6.51 and 7.00 Å, respectively, equivalent with effective vdW diameters of 2.66, 2.74 and 2.94 Å, for the Cl, Br and I halogen atoms in the $Nb_3X_8$ monolayers, respectively. Plotted stress-strain curves in Fig. 4 are uniaxial, which means that during the complete deformation and after the geometry minimization, these kagome monolayers exhibit stress component only along the loading direction and show negligible values along the two other perpendicular directions. As expected and stemmed from symmetrical hexagonal crystals, the stress-strain curves along the armchair and zigzag directions coincide for the initial linear sections, confirming isotropic elasticity in these nanosheets. According to conducted spin-polarized DFT calculations, the $C_{11}(C_{12})$ of the $Nb_3Cl_8$, $Nb_3Br_8$ and $Nb_3I_8$ monolayers are predicted to be 105 (27), 95 (23) and 81(18) GPa, respectively, corresponding to elastic modulus values of 98, 89 and 77 GPa, respectively, which are by less than 8% lower than the values predicted earlier [15].

Despite of exhibiting isotropic elasticity, results shown in Fig. 4 reveal anisotropic tensile behavior along these novel 2D systems, in which they yield higher tensile strengths along the zigzag direction than the armchair counterpart. The ultimate tensile strength of the $Nb_3Cl_8$, $Nb_3Br_8$ and $Nb_3I_8$ monolayers along the zigzag (armchair) directions are predicted to be 10.2 (8.1), 8.1 (7.0) and 5.9 (5.6) GPa, respectively. These results show a clear decreasing trend in the elastic modulus and tensile strength of the $Nb_3X_8$ nanosheets with the increase in the atomic number of halogen atoms. To better understand the reason for anisotropic tensile behavior in these kagome nanosheets, we plot the deformed $Nb_3Cl_8$ monolayer at strain levels after the ultimate tensile strength point in Fig. 4. We know that the orientation of bonds with respect to the loading play a critical role in the tensile strength values of a structure. As such, while bonds aligned toward the loading stretch and involve in the load transfer, those oriented perpendicular of loading contract and contribute marginally to the tensile strength. It appears that for the loading along the zigzag direction, because of the bonding architecture in these systems, more bonds stretch and involve in the loading, than the armchair direction (consider Fig. 4e and 4f). As shown in Fig. 4f, for the uniaxial loading along the armchair direction, one Nb-X in every unit cell elongates more considerably than other bonds in these systems, whereas along the zigzag direction more bonds uniformly contribute to the deformation. Our analysis of thickness evolution during the deformation, also revealed that in these systems, for the uniaxial loading along the armchair, the thickness decreases more considerably than along the zigzag. The reduction of thickness, although decreases the real volume of the system, but also facilitates the elongation of bonds along the in-plane direction, and overall results in a lower tensile strength. Results shown in Fig. 4 reveal that along the armchair direction, a clear bond breakage occurs in these system (see Fig. 4f), which is not obvious for the loading along the zigzag direction. It can be thus concluded that along the armchair direction these systems show more brittle failure behavior than the zigzag direction. This finding could be also realized from the stress-strain curves, because sharper drops in the stress values after reaching the ultimate tensile strength point are observable along the armchair direction.

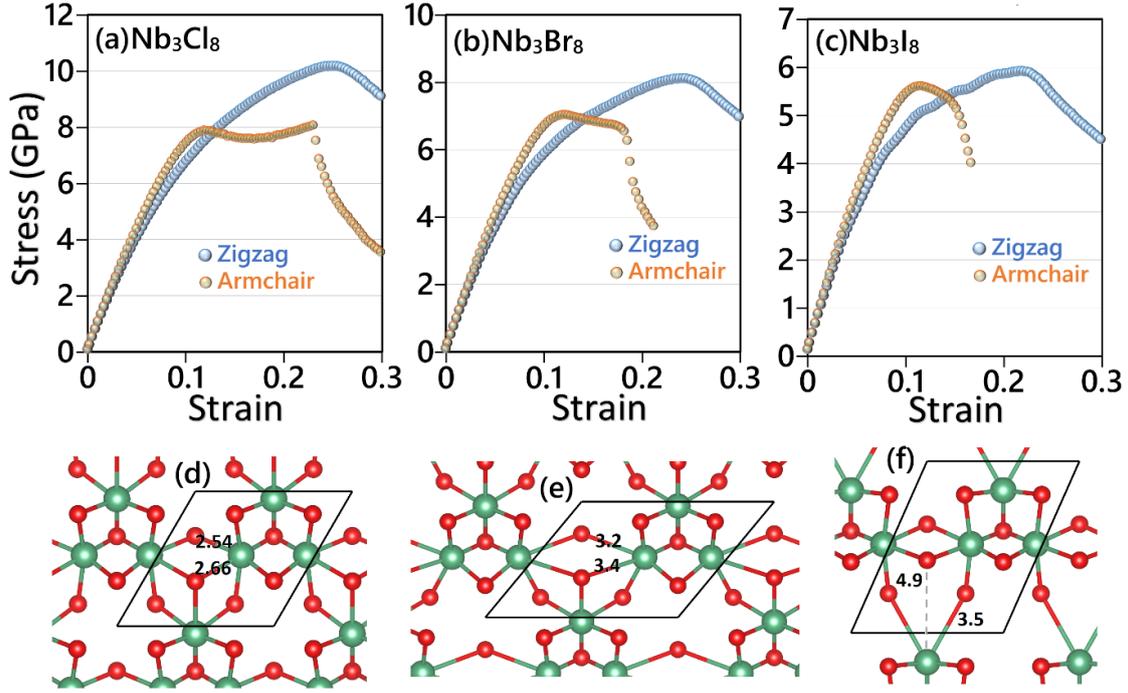

**Fig. 4**, True uniaxial stress-strain relations of the (a) $Nb_3Cl_8$, (b) $Nb_3Br_8$ and (c) $Nb_3I_8$ kagome monolayers. elongated along the armchair and zigzag directions. (d) Stress-free, (e) under-stress along the zigzag direction and (e) under-stress along the armchair direction single-layer $Nb_3Cl_8$ (values present bond lengths in Å).

## 4. Concluding remarks

Motivated by the successful exfoliation of the single-layer $Nb_3Cl_8$ kagome lattice with exciting electronic features [8], we herein presented spin-polarized DFT results on the physical properties of the $Nb_3X_8$ (X=Cl, Br, I) nanosheets. The exfoliation energies of 0.24, 0.27 and 0.28 J/m$^2$ are predicted for the $Nb_3Cl_8$, $Nb_3Br_8$ and $Nb_3I_8$ monolayers' isolation, which reveal that the separate layers in these systems show weak interactions, and confirm highly bright prospect for the experimental isolation of the $Nb_3Br_8$ and $Nb_3I_8$ monolayers from their native bulk structures. The predicted phonon dispersions confirm desirable dynamical stability of the $Nb_3Cl_8$, $Nb_3Br_8$ and $Nb_3I_8$ monolayers. By increasing the atomic number of halogen atoms in $Nb_3X_8$ systems, it is found that they show lower phonon group velocity, elastic modulus and tensile strength as well. The elastic modulus of the $Nb_3Cl_8$, $Nb_3Br_8$ and $Nb_3I_8$ monolayers are predicted to be 98, 89 and 77 GPa, respectively, which are independent of the loading direction. The ultimate tensile strength of the $Nb_3Cl_8$, $Nb_3Br_8$ and $Nb_3I_8$ monolayers along the zigzag (armchair) directions are predicted to be 10.2(8.1), 8.1 (7.0) and 5.9 (5.6) GPa, respectively. It is also found that along the armchair direction these systems show a more brittle failure behavior than the zigzag direction. Presented spin-polarized DFT results provide a brief but also important vision concerning the structural, dynamical stability, exfoliation energies and mechanical characteristics of the $Nb_3X_8$ (X=Cl, Br, I) nanosheets, which can serve as valuable information for their practical application in nanodevices

**Acknowledgment**

B.M. and X.Z. appreciate the funding by the Deutsche Forschungsgemeinschaft (DFG, German Research Foundation) under Germany's Excellence Strategy within the Cluster of Excellence PhoenixD (EXC 2122, Project ID 390833453). B. M and T. R. are greatly thankful to the VEGAS cluster at Bauhaus University of Weimar for providing the computational resources.

**Declarations Conflicts of interests**

The authors have no conflicts of interest to declare that are relevant to the content of this article.

**Appendix A. Supplementary data**

The following are the supplementary data to this article:

Nb₃Cl₈-1L
1.00000000000000
     6.7829799920112803    0.0000000000000000    0.0000000000000000
    -3.3914899959582780    5.8742329864630234    0.0000000000000000
    -0.000000000000000     0.0000000000000000   20.0000000000000000
   Nb   Cl
    3    8
Direct
  0.0561800853339968  0.5280900706669936  0.5178211720394165
  0.4719099593330017  0.5280900706669936  0.5178211720394165
  0.4719099593330017  0.9438199186660033  0.5178211720394165
  0.1683760807033427  0.3367521624066819  0.4351463289913037
  0.6632478075933228  0.8316239042966596  0.4351463289913037
  0.1683760957033404  0.8316239042966596  0.4351463289913037
  0.3333333429999996  0.6666666870000029  0.6101604919872622
  0.6666666870000029  0.3333333429999996  0.4496494500436901
  0.8347776837333417  0.1652223312666561  0.5859066222922918
  0.3304446635333087  0.1652223312666561  0.5859066222922918
  0.8347776837333417  0.6695553074666853  0.5859066222922918

Nb₃Cl₈-Bulk
   1.00000000000000
     6.7808874339922838   -0.0000000000004798    0.0000000000000000
    -3.3904437169469279    5.8724207780601159    0.0000000000000000
    -0.0000000000000000   -0.0000000000000000   12.3197509165851038
   Nb   Cl
    6    16
Direct
  0.0560976334521839  0.5280488407260863  0.7539405410432504
  0.4719511892739081  0.5280488407260875  0.7539405410432504
  0.4719511892739088  0.9439023785478166  0.7539405410432504
  0.5280488407260873  0.4719511892739078  0.2460594729567506
  0.9439023785478168  0.4719511892739088  0.2460594729567506
  0.5280488407260863  0.0560976334521841  0.2460594729567506
  0.1681442440056849  0.3362884890113612  0.6203493667225567
  0.6637114809886437  0.8318557409943175  0.6203493667225567
  0.1681442440056864  0.8318557409943158  0.6203493667225567
  0.8318557409943175  0.6637114809886435  0.3796506032774410
  0.8318557409943160  0.1681442440056866  0.3796506032774410
  0.3362884890113608  0.1681442440056849  0.3796506032774410
  0.6666666870000029  0.3333333429999996  0.0972537320656699
  0.3333333429999996  0.6666666870000029  0.9027462899343323
  0.3333333429999996  0.6666666870000029  0.3561696165735036
  0.6666666870000029  0.3333333429999996  0.6438303834264962
  0.6712992112738208  0.8356496356369097  0.1360916851646623
  0.1643503793630889  0.8356496356369086  0.1360916851646623
  0.1643503793630880  0.3287007597261737  0.1360916851646623
  0.8356496356369087  0.1643503793630891  0.8639083298353424
  0.3287007597261734  0.1643503793630880  0.8639083298353424
  0.8356496356369097  0.6712992112738206  0.8639083298353424

**Nb$_3$Br$_8$-1L**
```
1.00000000000000
     7.1208554645944444    0.0000000000000000    0.0000000000000000
    -3.5604277322339479    6.1668417290399358    0.0000000000000000
     0.0000000000000001    0.0000000000000000   20.0000000000000000
   Nb   Br
    3    8
Direct
  0.0628202571045215  0.5314101565522561  0.5179462170059116
  0.4685898734477324  0.5314101565522632  0.5179462170059116
  0.4685898734477395  0.9371797468954790  0.5179462170059116
  0.1674728743813954  0.3349457497627873  0.4292994906576862
  0.6650542202372102  0.8325271106186071  0.4292994906576862
  0.1674728893813860  0.8325271106186142  0.4292994906576862
  0.3333333429999996  0.6666666870000029  0.6176036223761140
  0.6666666870000029  0.3333333429999996  0.4467386551011968
  0.8359422363825867  0.1640577786174109  0.5901176371772946
  0.3281155582348254  0.1640577786174109  0.5901176371772946
  0.8359422363825867  0.6718844127651690  0.5901176371772946
```

**Nb$_3$Br$_8$-Bulk**
```
1.00000000000000
     7.1187546449574262    0.0000000000024340   -0.0000000000000000
    -3.5593773224245009    6.1650223658633330    0.0000000000000000
    -0.0000000000000000    0.0000000000000000   13.0200056764300829
   Nb   Br
    6   16
Direct
  0.0627212489194073  0.5313606484597021  0.7542024591342137
  0.4686393815402932  0.5313606484597021  0.7542024591342137
  0.4686393815402932  0.9372787630805934  0.7542024591342137
  0.5313606484597021  0.4686393815402932  0.2457975548657880
  0.9372787630805934  0.4686393815402932  0.2457975548657880
  0.5313606484597021  0.0627212489194073  0.2457975548657880
  0.1673413314413674  0.3346826638827242  0.6187827043951600
  0.6653173061172810  0.8326586535586351  0.6187827043951600
  0.1673413314413674  0.8326586535586351  0.6187827043951600
  0.8326586535586351  0.6653173061172810  0.3812172656048378
  0.8326586535586351  0.1673413314413674  0.3812172656048378
  0.3346826638827242  0.1673413314413674  0.3812172656048378
  0.6666666870000029  0.3333333429999996  0.0939071074762429
  0.3333333429999996  0.6666666870000029  0.9060929145237585
  0.3333333429999996  0.6666666870000029  0.3548480301535358
  0.6666666870000029  0.3333333429999996  0.6451519698464639
  0.6741661574468909  0.8370831087234447  0.1356333283004931
  0.1629169062765532  0.8370831087234447  0.1356333283004931
  0.1629169062765532  0.3258338135531030  0.1356333283004931
  0.8370831087234447  0.1629169062765532  0.8643666866995116
  0.3258338135531030  0.1629169062765532  0.8643666866995116
  0.8370831087234447  0.6741661574468909  0.8643666866995116
```

Nb$_3$I$_8$-1L
```
 1.00000000000000
     7.6422729266393024    0.0000000000000000    0.0000000000000000
    -3.8211364633696228    6.6184024971288382    0.0000000000000000
     0.0000000000000000    0.0000000000000000   20.0000000000000000
   Nb   I
    3    8
Direct
  0.0700184351835256  0.5350092515917692  0.4514171520008227
  0.4649907784082333  0.5350092515917692  0.4514171520008227
  0.4649907784082333  0.9299815568164809  0.4514171520008227
  0.1674941952448970  0.3349883904897868  0.3555414630147671
  0.6650115795102106  0.8325057897551053  0.3555414630147671
  0.1674941952448970  0.8325057897551053  0.3555414630147671
  0.3333333429999996  0.6666666870000029  0.5586994024087124
  0.6666666870000029  0.3333333429999996  0.3780638081563178
  0.8363970432817078  0.1636029867182946  0.5283030041293909
  0.3272059734365892  0.1636029867182946  0.5283030041293909
  0.8363970432817078  0.6727939675634094  0.5283030041293909
```

Nb$_3$I$_8$-Bulk
```
 1.00000000000000
     7.6306587895810392    0.0000000000072599   -0.0000000000000000
    -3.8153293947283804    6.6083443594124427    0.0000000000000000
    -0.0000000000000001    0.0000000000000000   13.9982948695326499
   Nb   I
    6   16
Direct
  0.0698578501537641  0.5349289490768799  0.7550781485300549
  0.4650710809231143  0.5349289490768815  0.7550781485300549
  0.4650710809231156  0.9301421618462367  0.7550781485300549
  0.5349289490768813  0.4650710809231141  0.2449218654699462
  0.9301421618462370  0.4650710809231156  0.2449218654699462
  0.5349289490768799  0.0698578501537643  0.2449218654699462
  0.1674860051884313  0.3349720113768606  0.6188045286563912
  0.6650279586231443  0.8325139798115712  0.6188045286563912
  0.1674860051884325  0.8325139798115698  0.6188045286563912
  0.8325139798115712  0.6650279586231442  0.3811954413436061
  0.8325139798115699  0.1674860051884328  0.3811954413436061
  0.3349720113768603  0.1674860051884313  0.3811954413436061
  0.6666666870000029  0.3333333429999996  0.0925757216897691
  0.3333333429999996  0.6666666870000029  0.9074243003102327
  0.3333333429999996  0.6666666870000029  0.3496623870764776
  0.6666666870000029  0.3333333429999996  0.6503376129235225
  0.6756607560613545  0.8378304080306803  0.1359735509655893
  0.1621696069693183  0.8378304080306794  0.1359735509655893
  0.1621696069693176  0.3243392149386396  0.1359735509655893
  0.8378304080306794  0.1621696069693184  0.8640264640344152
  0.3243392149386395  0.1621696069693176  0.8640264640344152
  0.8378304080306803  0.6756607560613542  0.8640264640344152
```